\documentclass[aps,pra,reprint,amsmath,amssymb,groupedaddress,showpacs]{revtex4-1}
\pdfoutput=1         
\usepackage{graphicx}
\usepackage{float}
\usepackage{dcolumn} 
\usepackage{bm}      
\usepackage[bookmarks=false]{hyperref} 
\hypersetup{
    unicode=false,          
    pdftoolbar=true,        
    pdfmenubar=true,        
    pdffitwindow=true,      
    pdfstartview={},        
    pdftitle={Hyperspherical Parameterization of Unitary Matrices}, 
    pdfauthor={Samuel R. Hedemann},
    pdfsubject={Unitary Matrices},   
    pdfcreator={PDFLaTeX},  
    pdfproducer={PDFLaTeX}, 
    pdfkeywords={Unitary,} {Matrix,} {Quantum,} {Hypersphere,} {Cayley-Klein,} {Gram-Schmidt,} {Interferometer,} {Entanglement}, 
    pdfnewwindow=true,      
    colorlinks=true,        
    linkcolor=black,        
    citecolor=black,        
    filecolor=black,        
    urlcolor=black          
}
\begin{document}
\title{Hyperspherical Parameterization of Unitary Matrices}
\author{Samuel R. Hedemann}
\affiliation{Dept.~of Physics and Engineering Physics, Stevens Institute of Technology, Hoboken, NJ 07030, USA}
\date{\today}
\begin{abstract}
Unitary operators are essential to quantum mechanics, however for discrete systems larger than a qubit, it is difficult to express them in a self-contained way.  This report presents just such a description, providing a compact, useful parameterization, with examples of physical applications.
\end{abstract}
\maketitle
\section{\label{sec:I}Introduction}
A unitary operator $U$ is defined by $U^{\dag}=U^{-1}$, meaning that its adjoint (Hermitian conjugate) is equal to its inverse, implying that $U^{\dag}U=UU^{\dag}=I$, where $I$ is the identity operator.  For discrete systems, unitary operators can be represented as matrices.  In general, the determinant of a unitary matrix is a unit complex number, and thus they are classified as unimodular, meaning $|\det(U)|=1$, since the modulus of the determinant is of unit length.  An important sub-group of the $n$-dimensional unitary group $\text{U}(n)$ is the special unitary group $\text{SU}(n)$, where ``special'' means $\det(U)=1$.

Unitary operators have many uses in quantum mechanics. First and foremost, they describe all transformations that simultaneously preserve state normalization, Hermiticity, and non-negativity, thus preserving the physicality of quantum states, both pure and mixed.  Time evolution of closed-system quantum states is a unitary process \cite[]{Saku}.  In quantum computation, unitary operations describe quantum logic gates \cite[]{Feyn,Niel}.  In particle physics, subatomic particles can be represented using special unitary groups \cite[]{Gold,Grif,Gell,Neem}.  Originally, unitary matrices were studied to provide alternative descriptions for the complicated rotational motions of gyroscopes, giving rise to the well-known Cayley-Klein (CK) parameterization of rotations using complex numbers \cite[]{Gold,Saku}.

A noteworthy feature of unitary operators is that any product of unitary operators is also unitary.  A related and useful property is that any $n$-dimensional unitary matrix can be factored into a product of single-qubit unitary matrices \cite[]{Niel}, which provides a means for realizing any unitary operation experimentally, as shown in \cite[]{Reck}. Due to their importance, there is a great wealth of information about unitary matrices, so in the present work we limit ourselves to only the bare essentials for our purpose.

The goal of this paper is to present a canonical form for unitary matrices in any dimension, subject to the constraints that it be as simple as possible without sacrificing generality, and that it be readily useful for common applications.  First, we give a general prescription in terms of CK parameters and constraints, then we incorporate the constraints through hyperspherical parameterization.  Finally, we examine several physical applications.
\section{\label{sec:II}Cayley-Klein Parameterization of Special Unitary Matrices}
Before proceeding, we must establish some practical conventions.  First, let our canonical form for $U$ have only single terms in its first row and first column, where the top row and the left column are both first.  The motivations for this are as follows.  As a change of basis, $U$ transforms the first basis element $|1\rangle$ to a vector $|1'\rangle$ with coefficients equal to the first row of $U$, such as
\begin{equation}
\left( {\begin{array}{*{20}c}
   {U_{1,1} } & {U_{1,2} }  \\
   {U_{2,1} } & {U_{2,2} }  \\
\end{array}}\! \right)\!\left( {\begin{array}{*{20}c}
   {|1\rangle }  \\
   {|2\rangle }  \\
\end{array}}\! \right) = \left( {\begin{array}{*{20}c}
   {U_{1,1} |1\rangle  + U_{1,2} |2\rangle }  \\
   {U_{2,1} |1\rangle  + U_{2,2} |2\rangle }  \\
\end{array}}\! \right)\equiv \left( {\begin{array}{*{20}c}
   {|1'\rangle }  \\
   {|2'\rangle }  \\
\end{array}}\! \right)\!.
\label{eq:1}
\end{equation}
Additionally, if a pure state in the first basis state $|\psi\rangle=|1\rangle$ is operated on by $U$, it is transformed into a state whose coefficients are the first column of $U$, such as
\begin{equation}
U|1\rangle  = \left( {\begin{array}{*{20}c}
   {U_{1,1} } & {U_{1,2} }  \\
   {U_{2,1} } & {U_{2,2} }  \\
\end{array}} \right)\left( {\begin{array}{*{20}c}
   1  \\
   0  \\
\end{array}} \right) = \left( {\begin{array}{*{20}c}
   {U_{1,1} }  \\
   {U_{2,1} }  \\
\end{array}} \right).
\label{eq:2}
\end{equation}
The origin of this first convention is two-fold; in general it is only possible to get at most one row and one column of a unitary matrix to have single terms in all entries, and focusing on the first row and first column vectors (leading vectors) affords simple referencing in equations.  

As a second convention to minimize notation, for $n\geq 3$, let the first row and first column of $U$ be free of negative signs in a CK parameterization, and let the top row be free of explicit complex conjugate symbols.
 
For simplicity, let $U\in\text{SU}(n)$ be \textit{special} unitary in this section, since given a special unitary matrix we need only multiply by a unit complex number to get a general unitary matrix, and other sub-groups can be reached via any additional general unitary transformations.  

Letting $n$ be the dimension, there are two special cases.  If $n=1$, the only special unitary matrix is the scalar,
\begin{equation}
U^{[1]}=1,
\label{eq:3}
\end{equation}
where numbers in square brackets of superscripts denote dimension $n$.  For $n=2$, $U$ is not unique, and we have
\begin{equation}
U^{[2]}  = \left( {\begin{array}{*{20}c}
   a & b  \\
   { - b^* } & {a^* }  \\
\end{array}} \right),
\label{eq:4}
\end{equation}
where $a$ and $b$ are complex numbers where $|a|^2+|b|^2 =1$.
\subsection{\label{subsec:II-A}General Formula for $n\geq 1$}
For all dimensions, $n\geq 1$, $U^{[n]}$ is given by
\begin{equation}
U^{[n]}  = \Phi ^{[n]} \prod\limits_{\alpha  = 2}^n {\prod\limits_{\beta  = 1}^{\alpha  - 1} {\Omega _{(\alpha ,\beta )}^{[n]} ( - a_{\alpha ,\beta }^* ,b_{\alpha ,\beta }^* )} },
\label{eq:5}
\end{equation}
where products are written left to right, parenthetical subscripts on matrices indicate labels, not matrix elements, and where $\Omega _{(\alpha ,\beta )}^{[n]}(x,y)$ is the qubit factor matrix,
\begin{equation}
\Omega _{(\alpha ,\beta )}^{[n]} (x,y)\! \equiv I_{(\alpha ,\beta )}^{[n]}  + \left\{\!\! {\begin{array}{*{20}l}
   {M_{(\alpha ,\beta )}^{[n]} (x,y)}; &\!\! {\alpha\!  +\! \beta  \ne 3+\! \delta_{n,2}}  \\
   {M_{(\alpha ,\beta )}^{[n]} (x,\!\gamma _n y^* )}; &\!\! {\alpha\!  +\! \beta  = 3+\! \delta_{n,2}},  \\
\end{array}} \right.
\label{eq:6}
\end{equation}
where $\delta_{j,k}$ is the Kronecker delta and $I_{(\alpha ,\beta )}^{[n]}$ is the subspace identity matrix,
\begin{equation}
I_{(\alpha ,\beta )}^{[n]}  \equiv \sum\limits_{\scriptstyle m = 1 \hfill \atop 
  \scriptstyle m \ne \alpha ,\beta  \hfill}^n \!\!{E_{(m,m)}^{[n]} },
\label{eq:7}
\end{equation}
in which $E_{(a,b)}^{[n]}\equiv |a\rangle\langle b|$ is the $n$-dimensional elementary matrix with a $1$ in the row-a, column-b entry, and $0$ elsewhere.  The matrix $M_{(\alpha ,\beta )}^{[n]} (s,t)$ from (\ref{eq:6}) is given by
\begin{equation}
M_{(\alpha ,\beta )}^{[n]} (s,t) \equiv \begin{array}{l}
 {\kern 8pt}Q_{1,1} (s,t)E_{(\beta ,\beta )}^{[n]}  + Q_{1,2} (s,t)E_{(\beta ,\alpha )}^{[n]}  \\ 
  + Q_{2,1} (s,t)E_{(\alpha ,\beta )}^{[n]}  + Q_{2,2} (s,t)E_{(\alpha ,\alpha )}^{[n]},  \\ 
 \end{array}
\label{eq:8}
\end{equation}
where $\alpha >\beta$, and $Q(u_1,u_2)$ is the single qubit matrix,
\begin{equation}
Q(u_1 ,u_2 ) \equiv \left( {\begin{array}{*{20}c}
   {u_2 } & {u_1 }  \\
   { - u_1^* } & {u_2^* }  \\
\end{array}} \right),
\label{eq:9}
\end{equation}
which has its alphabet of ordered variables reversed from that of (\ref{eq:4}).  The factor of $\gamma_n$ in (\ref{eq:6}) is defined by
\begin{equation}
\gamma _n  \equiv ( - 1)^{\frac{1}{4}({2n - 1 + ( - 1)^n })}.
\label{eq:10}
\end{equation}
The matrix $\Phi^{[n]}$ from (\ref{eq:5}) is the modified flip-matrix,
\begin{equation}
\Phi ^{[n]}  \equiv \gamma _n E_{(n,1)}^{[n]}  + \sum\limits_{m = 1}^{n-1} {E_{(m,n - m + 1)}^{[n]} }.
\label{eq:11}
\end{equation}
If $\gamma _n\equiv 1$ in (\ref{eq:11}), $\Phi ^{[n]}(\gamma_{n}\equiv 1)$ would oscillate between special and anti-special, such that starting at $n=1$, $\det(\Phi^{[n]}(\gamma_{n}\equiv 1))=\{1,-1,-1,1,1,-1,-1,\ldots\}$, which is precisely the sequence generated by $\gamma _n$.  Thus $\gamma _n$ ensures that $\Phi ^{[n]}$ is always special without introducing any complex factors.  The complex number pairs $a_{\alpha,\beta}$ and $b_{\alpha,\beta}$ from (\ref{eq:5}) are CK pairs, each subject to the constraint that $|a_{\alpha,\beta}|^2+|b_{\alpha,\beta}|^2 =1$.  There are $\frac{n^2 -n}{2}$ such pairs of parameters, with each pair corresponding to a special unitary transformation on a particular single qubit subspace, one corresponding to each lower-off-diagonal element, which are used to index the qubit rotations.
\subsection{\label{subsec:II-B}Examples for Cayley-Klein Form}
To demonstrate the use of (\ref{eq:5}), for $n=3$, if we let
\begin{equation}
\begin{array}{*{20}c}
   {(a_{2,1} ,b_{2,1} ) \equiv (e,f)} & {}  \\
   {(a_{3,1} ,b_{3,1} ) \equiv (a,b)} & {(a_{3,2} ,b_{3,2} ) \equiv (c,d)},  \\
\end{array}
\label{eq:12}
\end{equation}
then (\ref{eq:5}) produces
\begin{equation}
U^{[3]}  = \left( {\begin{array}{*{20}c}
   a & {bc} & {bd}  \\
   {b^* e} & { - a^* ce - d^* f^* } & { - a^* de + c^* f^* }  \\
   {b^* f} & { - a^* cf + d^* e^* } & { - a^* df - c^* e^* }  \\
\end{array}} \right),
\label{eq:13}
\end{equation}
subject to the constraints $|a|^2+|b|^2 =1$, $|c|^2+|d|^2 =1$, and $|e|^2+|f|^2 =1$.  Notice that the first row and column conform to our conventions, and that in abbreviating the parameters in (\ref{eq:12}), we started with those corresponding to the bottom left corner, and then moved right along the bottom row, and then moved up the first column.  For $n=4$, if we define the abbreviations
\begin{equation}
\begin{array}{*{20}l}
   {(a_{2,1} ,b_{2,1} )}  \\
   {(a_{3,1} ,b_{3,1} )}  \\
   {(a_{4,1} ,b_{4,1} )}  \\
\end{array}\!\!\!\begin{array}{*{20}l}
   { \equiv\! (j,k)}  \\
   { \equiv\! (g,h)}  \\
   { \equiv\! (a,b)}  \\
\end{array}\!\begin{array}{*{20}l}
   {}  \\
   {(a_{3,2} ,b_{3,2} )}  \\
   {(a_{4,2} ,b_{4,2} )}  \\
\end{array}\!\!\!\begin{array}{*{20}l}
   {}  \\
   { \equiv\! (l,m)}  \\
   { \equiv\! (c,d)}  \\
\end{array}\!\!\begin{array}{*{20}l}
   {}  \\
   {}  \\
   {(a_{4,3} ,b_{4,3} )}  \\
\end{array}\!\!\! \begin{array}{*{20}l}
   {}  \\
   {}  \\
   { \equiv\! (e,f)},  \\
\end{array}
\label{eq:14}
\end{equation}
then, from (\ref{eq:5}) we obtain
\begin{equation}
\begin{array}{l}
 U^{[4]}  =  \\ 
 \left(\!\! {\begin{array}{*{20}c}
   a & {bc} & {bde} & {bdf}  \\
   {b^* g} & {\left(\!\! \begin{array}{l}
  - a^* cg \\ 
  + d^* hl \\ 
 \end{array}\!\! \right)} & {\left(\!\! \begin{array}{l}
  - a^* deg  \\ 
  - c^* ehl \\ 
  + f^* hm \\ 
 \end{array}\!\! \right)} & {\left(\!\! \begin{array}{l}
  - a^* dfg \\ 
  - c^* fhl \\ 
  - e^* hm \\ 
 \end{array}\!\! \right)}  \\
   {b^* h^* j} & {\left(\!\! \begin{array}{l}
  - a^* ch^* j \\ 
  - d^* g^* jl \\ 
  + d^* k^* m^*  \\ 
 \end{array}\!\! \right)} & {\left(\!\! \begin{array}{l}
  - a^* deh^* j \\ 
  + c^* eg^* jl \\ 
  - c^* ek^* m^*  \\ 
  - f^* g^* jm \\ 
  - f^* k^* l^*  \\ 
 \end{array}\!\!\! \right)} & {\left(\!\! \begin{array}{l}
  - a^* dfh^* j \\ 
  + c^* fg^* jl \\ 
  - c^* fk^* m^*  \\ 
  + e^* g^* jm \\ 
  + e^* k^* l^*  \\ 
 \end{array}\!\!\! \right)}  \\
   {b^* h^* k} & {\left(\!\! \begin{array}{l}
  - a^* ch^* k \\ 
  - d^* g^* kl \\ 
  - d^* j^* m^*  \\ 
 \end{array}\!\! \right)} & {\left(\!\! \begin{array}{l}
  - a^* deh^* k \\ 
  + c^* eg^* kl \\ 
  + c^* ej^* m^*  \\ 
  - f^* g^* km \\ 
  + f^* j^* l^*  \\ 
 \end{array}\!\!\! \right)} & {\left(\!\! \begin{array}{l}
  - a^* dfh^* k \\ 
  + c^* fg^* kl \\ 
  + c^* fj^* m^*  \\ 
  + e^* g^* km \\ 
  - e^* j^* l^*  \\ 
 \end{array}\!\!\! \right)}  \\
\end{array}}\!\!\! \right) \\ 
 \end{array}\!\!,
\label{eq:15}
\end{equation}
subject to $|a|^2 +|b|^2 =1$, $|c|^2 +|d|^2 =1$, $|e|^2 +|f|^2 =1$, $|g|^2 +|h|^2 =1$, $|j|^2 +|k|^2 =1$, and $|l|^2 +|m|^2 =1$.  Symbolic and numerical tests up to $n=7$ verify that (\ref{eq:5}) is special unitary and conforms to all proposed conventions.

To reposition the single-term vectors of $U^{[n]}$, use
\begin{equation}
U_{(r,c|r_{0},c_{0})}^{[n]}\! \equiv\! {{\det}^{\frac{1}{n}} ( {S_{(r_0 ,r)}^{[n]} S_{(c_0 ,c)}^{[n]} } )} S_{(r_0 ,r)}^{[n]} U_{(r_{0},c_{0}|r_{0},c_{0})}^{[n]} S_{(c_0 ,c)}^{[n]},
\label{eq:16}
\end{equation}
where $(r,c)$ are the row and column to which you want the single-term vectors to move from $(r_0,c_0)$, and
\begin{equation}
S_{(z_1 ,z_2 )}^{[n]}\!  \equiv (1\! - \delta _{z_1 ,z_2 } )E_{(z_1 ,z_2 )}^{[n]}\!  + E_{(z_2 ,z_1 )}^{[n]}\!  +I_{(z_{1},z_{2})}^{[n]}
\label{eq:17}
\end{equation}
is a swap matrix.  The scalar in (\ref{eq:16}) keeps it special but adds an explicit phase factor which can be removed from the single-term vectors by redefining their Cayley-Klein pairs as needed.  Note that (\ref{eq:5}) produces $U^{[n]}\equiv U_{(1,1|1,1)}^{[n]}$.
\section{\label{sec:III}Hyperspherical Parameterization of Special Unitary Matrices}
The Cayley-Klein parameterization convention developed in the previous section gives us a methodical and orderly framework for describing special unitary matrices.  Here, we adopt additional conventions which ensure a clear correspondence between the leading state vectors of $U$ and a hyperspherical representation of a pure state.
\subsection{\label{subsec:III-A}Review of Hyperspherical Coordinates}
An $(n-1)$-sphere is the surface of all points equidistant from a single point in $n$-dimensional Euclidean space ($n$-space).  Therefore, in 3-space the 2-sphere is an ordinary sphere.  An $(n-1)$-sphere can be parameterized by a single radial coordinate $r$, and $n-1+\delta_{n,1}$ angular coordinates, $n-2+\delta_{n,1}$ of which are polar coordinates each with range $[0,\pi]$, and one of which is azimuthal with range $[0,2\pi)$ if $n\geq 2$, (or with two-point range $\{0,\pi\}$ if $n=1$), as shown for $n\geq 2$ below \cite[]{Ratc},
\begin{equation}
\begin{array}{*{20}l}
 {\text{Radial:}}   & {\;\;\;\;{\kern 1pt}r \in[0,\infty)} \\ 
 {\text{Polar:}}    & {\left\{ \begin{array}{l}
                       \theta_1 \in[0,\pi] \\ 
                       \vdots  \\ 
                       \theta_{n-2} \in[0,\pi] \\ 
                       \end{array} \right.
                    } \\ 
 {\text{Azimuthal:}}& {\;\;\;\;{\kern 1pt}\theta_{n-1} \in[0,2\pi).} \\ 
\end{array}
\label{eq:18}
\end{equation}

For $n\geq 1$, a formula to express the $n$ Cartesian coordinates of $n$-space using hyperspherical coordinates is
\begin{equation}
x_{ \scriptstyle 1\leq k \leq n-1+\delta_{n,1} \hfill \atop 
  \scriptstyle k=n\neq 1 \hfill }(r,\{\theta_u\})  = r\left({\prod\limits_{m = 1}^{{\scriptstyle k - 1 \hfill \atop 
  \scriptstyle k - 2 \hfill} } {\sin (\theta _m )} }\right)  \begin{array}{l}
 \cos (\theta _k ) \\ 
 \sin (\theta _{k - 1} ) \\ 
 \end{array},
\label{eq:19}
\end{equation}
where the top row is the $k^{\text{th}}$ Cartesian coordinate where $k\neq n$ unless $n=1$, and the bottom row is the $n^{\text{th}}$ Cartesian coordinate when $n\neq 1$, empty products are $1$, and $\{\theta_{u}\}\equiv \{\theta_{1},\ldots ,\theta_{n-1+\delta_{n,1}}\} $.  Inverses are
\begin{equation}
\begin{array}{l}
 r = \sqrt{\sum\limits_{m=1}^{n}\! {x_m^2} }{\kern 1pt},\;\; \theta_{1\leq k\leq n-2}=\cot^{-1}\!\left({{x_{k} \mathord{\left/
 {\vphantom {x_{k} {\!\sqrt{\sum\limits_{m=k+1}^n\!\! {x_m^2}}}}} \right.
 \kern-\nulldelimiterspace} {\!\sqrt{\sum\limits_{m=k+1}^n\!\! {x_m^2}}}}}\,\right)\!, \\
 \cos(\theta_{n-1})=\frac{x_{n-1}^{\;}}{\sqrt{x_{n-1}^{2}+x_{n}^{2}}},\;\;\sin(\theta_{n-1})=\frac{x_{n}}{\sqrt{x_{n-1}^{2}+x_{n}^{2}}},\\
\end{array}
\label{eq:20}
\end{equation}
valid for $n\geq 3$ for simplicity, though a more general form exists, and the azimuthal angle is left in its sine and cosine forms since both are needed to find $\theta_{n-1}$.
\subsection{\label{subsec:III-B}Representing Pure States on Hyperspheres}
A pure state with global phase has the form
\begin{equation}
|\psi \rangle  = \sum\limits_{k = 1}^n {r_k e^{i\phi _{k} } |k\rangle },
\label{eq:21}
\end{equation}
with real moduli $r_{1\leq k \leq n}\in [0,1]$, real phase angles $\phi_{1\leq k \leq n}\in [0,2\pi)$, and kets $|1\leq k \leq n\rangle$ forming a complete basis in a Hilbert space of $n$ dimensions, and
\begin{equation}
\sum\limits_{k = 1}^n {r_k^2 }  = 1,
\label{eq:22}
\end{equation}
which suggests that the (non-negative) $r_k$ can be described in  the first sector of a unit hypersphere in $n$-space, while the $\phi_k$ can be represented on $n$ separate circles.  Thus, for $n\geq 1$, the $\{r_{k}\}\equiv \{r_{1},\ldots ,r_{n}\}$ can be parameterized as
\begin{equation}
r_{k}\equiv x_{k}(1,\{\theta_{u}\});\;\;\theta _u  \in (1 - \delta _{n,1} )[0,{\textstyle{\pi  \over 2}}],
\label{eq:23}
\end{equation}
where $x_k(r,\{ \theta_u \})$ is given in (\ref{eq:19}), and each phase angle $\phi_k$ is already its own parameterization.  The mapping of the $r_k$ in (\ref{eq:23}) only uses the \textit{first sector} of a unit hypersphere because remapping to a full hypersphere does not give unique locations to all basis states.  This method is in contrast to the Bloch sphere which uses relative phase together with the polar angle to put all coordinates into a \textit{single} object.  Here, the geometrical interpretation is that the superposition properties of $|\psi\rangle$ are represented on a hypersphere, while the absolute phase of each outcome is represented on a separate \textit{phase circle}, independent of the \textit{superposition hypersphere}. 

Inverse equations for pure states when $n\geq 1$ are
\begin{equation}
\begin{array}{l}
 \theta _{1 \le k \le n - 1+\delta_{n,1}}  = \cot ^{ - 1}\! \left( {{{|c_k |\!} \mathord{\left/
 {\vphantom {{|c_k |\!} {\!\!\sqrt {\sum\limits_{m = k + 1}^n \!\!{|c_m |^2 } } }}} \right.
 \kern-\nulldelimiterspace} {\!\!\sqrt {\sum\limits_{m = k + 1}^n \!\!{|c_m |^2 } } }}}\, \right) \\ 
 \phi _{1 \le k \le n}  = \arg (c_{k} ), \\ 
 \end{array}
\label{eq:24}
\end{equation}
where $c_k \equiv \langle k|\psi\rangle \equiv r_{k}e^{i\phi_{k}}$, and $\arg(re^{i\phi})\equiv \phi\in [0,2\pi)$.

An example of (\ref{eq:23}) applied to (\ref{eq:21}) for $n=4$ is
\begin{equation}
|\psi \rangle  = \begin{array}{*{20}l}
   {\;\;\;\cos (\theta_{1})e^{i\phi _1 }|1\rangle }  \\
   { + \sin (\theta_{1})\cos (\theta_{2})e^{i\phi _2 } |2\rangle }  \\
   { + \sin (\theta_{1})\sin (\theta_{2})\cos (\theta_{3})e^{i\phi _3 } |3\rangle }  \\
   { + \sin (\theta_{1})\sin (\theta_{2})\sin (\theta_{3})e^{i\phi _4 } |4\rangle },  \\
\end{array}
\label{eq:25}
\end{equation}
where $\theta_{1},\theta_{2},\theta_{3}\in [0,\frac{\pi}{2}]$ and $\phi_{1},\phi_{2},\phi_{3},\phi_{4}\in [0,2\pi)$.  Thus, the superposition properties of this state are portrayed by three angles $\theta_k$ in the first sector of a unit hypersphere in $4$-space, while there are four separate absolute phase circles portrayed by the $\phi_k$, or if global phase can be discarded, just three relative phase circles portrayed by $\phi_{2|1}\equiv\phi_{2}-\phi_{1}$, $\phi_{3|1}\equiv\phi_{3}-\phi_{1}$, and $\phi_{4|1}\equiv\phi_{4}-\phi_{1}$.

In general, a pure state has one superposition hypersphere and $n-1$ relative phase circles, resulting in $n$ separate geometric objects to describe a pure state in $n$-dimensional Hilbert space.  An alternative geometrical approach is to use $2n-1$ Cartesian coordinates for the complex parts of (\ref{eq:21}) without global phase to represent a pure state on a \textit{single} hypersphere in $(2n-1)$-space.  However, that method obscures the physical roles of its parameters with respect to superposition and phase, while the method of (\ref{eq:23}) yields clear relationships between its parameters and physical properties.
\subsection{\label{subsec:III-C}Application to Unitary Matrices} 
At first glance, it would \textit{seem} that the most general way to build the CK constraints directly into $U^{[n]}$ is
\begin{equation}
(a_{j,k} ,b_{j,k} ) \equiv (c_{j,k} e^{i\phi _{j,k} } ,s_{j,k} e^{i\varepsilon _{j,k} } );\;\!\begin{array}{*{20}l}
   {2\leq j \leq n}  \\
   {1\leq k \leq j-1,}  \\
\end{array}
\label{eq:26}
\end{equation}
where $c_{j,k}\!\equiv\! \cos(\vartheta_{j,k})$, $s_{j,k}\!\equiv \! \sin(\vartheta_{j,k})$, and $\vartheta_{j,k},\phi_{j,k},\varepsilon_{j,k}\!\in\! [0,2\pi)$.  However, this introduces \textit{more} variables than the true number $D$ of degrees of freedom in $U^{[n]}$.  It turns out that there exists an \textit{essential} parameterization such that $D$ for special and general unitary matrices is
\begin{equation}
D_{\text{SU}(n)}=n^{2}-1 \;\;\text{and}\;\; D_{\text{U}(n)}=n^{2}.
\label{eq:27}
\end{equation}
Since the parameterization of (\ref{eq:26}) uses $3(\frac{n^2 -n}{2})$ variables, this means that it introduces $\frac{(n-1)(n-2)}{2}$ \textit{irrelevant} parameters, a waste of computation time that scales as $\frac{n^{2}}{2}$ for large $n$.  Therefore, to further constrain (\ref{eq:26}) to have only the \textit{essential} parameters, let
\begin{equation}
\varepsilon _{j,k}  \equiv \chi _{j,k}\delta _{j - 1,k},
\label{eq:28}
\end{equation}
where $\chi\in [0,2\pi)$.  Thus, (\ref{eq:28}) ensures that (\ref{eq:26}) only injects $n^2 -1$ variables into $U^{[n]}$.  The meaning of this is that two phase angles are needed by all single qubit rotations whose nontrivial off-diagonal elements are directly next to the main diagonal elements, while all other rotations only need one phase angle.  In terms of quantum states, this introduces $\frac{n^2 -n}{2}$ superposition angles and $\frac{n^2 +n}{2}-1$ phase angles for $\text{SU}(n)$.  To map the superposition angles onto first sectors of hyperspheres, use
\begin{equation}
\vartheta_{j,k}  \equiv \theta_{j,k}\in (1-\delta_{n,1})[0,{\textstyle{\pi  \over 2}}].
\label{eq:29}
\end{equation}
Thus (\ref{eq:29}) allows us to relate the form of $U^{[n]}$ to quantum pure states in a way particularly evident in its leading vectors.  To obtain a general unitary matrix $G^{[n]}\in\text{U}(n)$, simply apply the essential parameterization to $U^{[n]}$, and then let $G^{[n]}\equiv e^{i\gamma}U^{[n]}$, where $\gamma\in[0,2\pi)$.
\subsection{\label{subsec:III-D}Examples of Hyperspherical Unitary Matrices}
To illustrate the benefits of the above parameterizations, for $n=3$, using (\ref{eq:26}) and (\ref{eq:28}) in (\ref{eq:13}) yields
\begin{equation}
\begin{array}{l}
 U^{[3]}  =  \\ 
 \left(\!\!\! {\begin{array}{*{20}c}
   {F_{\!\vartheta _{3,1} ,0,0}^{\phi _{3,1}  + \pi ,0} } &\!\!\! {F_{\!\vartheta _{3,1}  - {\textstyle{\pi  \over 2}},\vartheta _{3,2} ,\pi }^{\phi _{3,2} ,0} } &\!\!\!\!\! {F_{\!\vartheta _{3,1}  - {\textstyle{\pi  \over 2}},\vartheta _{3,2}  - {\textstyle{\pi  \over 2}},\pi }^{\chi _{3,2} ,0} }  \\
   {F_{\!\vartheta _{3,1}  - {\textstyle{\pi  \over 2}},\pi ,\vartheta _{2,1} }^{\phi _{2,1} ,0} } &\!\!\! {F_{\!\vartheta _{3,1} ,\vartheta _{3,2} ,\vartheta _{2,1} }^{\!\left(\!\! {\scriptstyle  - \phi _{3,1}  + \phi _{3,2}  + \phi _{2,1} , \hfill \atop 
  \scriptstyle \chi _{3,2}  + \chi _{2,1}  \hfill}\!\! \right)} } &\!\!\!\!\! {F_{\!\vartheta _{3,1} ,\vartheta _{3,2}  - {\textstyle{\pi  \over 2}},\vartheta _{2,1} }^{\!\left(\!\! {\scriptstyle  - \phi _{3,1}  + \chi _{3,2}  + \phi _{2,1} , \hfill \atop 
  \scriptstyle \phi _{3,2}  + \chi _{2,1}  \hfill}\!\! \right)} }  \\
   {F_{\!\vartheta _{3,1}  -\! {\textstyle{\pi  \over 2}},\pi ,\vartheta _{2,1}  - {\textstyle{\pi  \over 2}}}^{\chi _{2,1} ,0} } &\!\!\! {F_{\!\vartheta _{3,1} ,\vartheta _{3,2} ,\vartheta _{2,1}  - {\textstyle{\pi  \over 2}}}^{\!\left(\!\! {\scriptstyle  - \phi _{3,1}  + \phi _{3,2}  + \chi _{2,1} , \hfill \atop 
  \scriptstyle \chi _{3,2}  + \phi _{2,1}  \hfill}\!\! \right)} } &\!\!\!\!\! {F_{\!\vartheta _{3,1} ,\vartheta _{3,2}  -\! {\textstyle{\pi  \over 2}},\vartheta _{2,1}\!  -\! {\textstyle{\pi  \over 2}}}^{\!\left(\!\!\! {\scriptstyle  - \phi _{3,1}  + \chi _{3,2}  + \chi _{2,1} , \hfill \atop 
  \scriptstyle \phi _{3,2}  + \phi _{2,1}  \hfill}\!\!\! \right)} }  \\
\end{array}}\!\!\!\! \right) \\ 
 \end{array}
\label{eq:30}
\end{equation}
where, given $c(x)\equiv \cos(x)$ and $s(x)\equiv \sin(x)$, we define
\begin{equation}
F_{v,w,x}^{y,z}  \equiv  - c(v)c(w)c(x)e^{iy}  - s(w)s(x)e^{ - iz}.
\label{eq:31}
\end{equation}
Thus, a single function can be used to describe all elements.  Employing the first-sector hypersphere mapping of (\ref{eq:29}) then reveals the columnized first row of (\ref{eq:30}) as
\begin{equation}
|U_{1,:}^{[3]} \rangle  \equiv \left( {\begin{array}{*{20}c}
   {U_{1,1}^{[3]} }  \\
   {U_{1,2}^{[3]} }  \\
   {U_{1,3}^{[3]} }  \\
\end{array}} \right) =\! \left(\! {\begin{array}{*{20}c}
   {c(\theta_{3,1})e^{i\phi_{3,1} } }  \\
   {s(\theta_{3,1})c(\theta_{3,2})e^{i\phi_{3,2} } }  \\
   {s(\theta_{3,1})s(\theta_{3,2})e^{i\chi_{3,2} } }  \\
\end{array}}\! \right),
\label{eq:32}
\end{equation}
where $\theta_{3,1},\theta_{3,2}\in [0,\frac{\pi}{2}]$, and $\phi_{3,1},\phi_{3,2},\chi_{3,2}\in [0,2\pi)$, so this vector has the standard hyperspherical form of a pure state, with a similar result holding for the first column of $U^{[3]}$, in keeping with our conventions.

For $n=4$, the first row of $U^{[4]}$ has the form of (\ref{eq:25}), and using the methods presented in this paper, this same pattern is maintained for all dimensions $n\geq 1$.  The true benefit of the form in (\ref{eq:30}), and of this method for all $n$, is that it \textit{incorporates} the CK constraints automatically, while using the fewest parameters possible.
\section{\label{sec:IV}Physical Applications}
While the conventions presented in this work are interesting mathematically, their worth to physicists will become apparent in the many ways their use can streamline and enhance the field of quantum mechanics.  To illustrate this, we now discuss a few physical applications.
\subsection{\label{subsec:IV-A}Complete Basis Generation}
Given a state vector $|\psi\rangle$ in a Hilbert space of dimension $n\geq 2$, suppose we wish to find a set of $n-1$ vectors that are orthonormal to it.  The standard prescription for this task is the Gram-Schmidt orthonormalization procedure, but it can be tedious to use and it requires that we start with a set of $n$ vectors.  

An interesting alternative arises from the methods of this paper.  First, note that for a unitary matrix constructed from (\ref{eq:5}), the leading vectors have single-term elements.  This yields simple expressions for the first $2n-1$ essential parameters of a unitary matrix by using the hyperspherical inverse equations of $|\psi\rangle$ in (\ref{eq:24}).  Then, we use the property that all the rows of a unitary matrix form a complete basis, obtaining our $n-1$ new vectors from the remaining rows of $U^{[n]}$ up to $n(n-2)$ undetermined degrees of freedom that can be freely chosen.

Therefore, to find the complete basis of states that includes the known vector,
\begin{equation}
|\psi \rangle  \equiv \sum\limits_{k = 1}^n {c_k |k\rangle },
\label{eq:33}
\end{equation}
first get the superposition and phase angles of (\ref{eq:33}) from (\ref{eq:24}).  Then, to relate them to individual qubit rotations using our conventions, use the ordering described in the Cayley-Klein examples to relabel these angles as the dual-indexed angles,
\begin{equation}
\begin{array}{*{20}r}
   {(\theta _{n,k} ,\phi _{n,k} )}  \\
   {\chi _{n,n - 1} }  \\
\end{array}\begin{array}{*{20}l}
   { \equiv (\theta _k ,\phi _k );\;\; 1\leq k\leq n - 1}  \\
   { \equiv \phi _n. }  \\
\end{array}
\label{eq:34}
\end{equation}
Thus, defining the CK pairs using (\ref{eq:26}), (\ref{eq:28}), and (\ref{eq:29}), all essential parameters whose first index is $n$ have values given by (\ref{eq:34}) which defines the determined CK pairs $(a_{n,1},b_{n,1}),(a_{n,2},b_{n,2}),\ldots,(a_{n,n-1},b_{n,n-1})$.

The remaining CK pairs can be freely chosen.  Specifically, we can arbitrarily choose all $(a_{j,k},b_{j,k})$ for $2\leq j\leq n-1$ and $1\leq k\leq j-1$ such that $|a_{j,k}|^2+|b_{j,k}|^2 =1$.  These constraints can be satisfied automatically with the fewest parameters by using the essential parameterization of (\ref{eq:26}), (\ref{eq:28}), and (\ref{eq:29}), which then requires only $n(n-2)$ free parameters to be chosen.

Then, putting the determined and freely chosen CK pairs into (\ref{eq:5}) produces a matrix $U^{[n]}$ where the first row vector is the generating state $|\psi\rangle$, and the remaining $n-1$ row vectors are orthonormal to $|\psi\rangle$ and each other.  Therefore, using the method just described, it is possible to generate a complete basis of states from a single input state $|\psi\rangle$ and $n(n-2)$ free real scalars.

For example, to find an orthonormal basis that includes
\begin{equation}
|\psi \rangle  = c_1 |1\rangle  + c_2 |2\rangle  + c_3 |3\rangle  + c_4 |4\rangle ,
\label{eq:35}
\end{equation}
first use (\ref{eq:24}) to get its hyperspherical coordinates as
\begin{equation}
\begin{array}{l}
 \theta _1  = \cot ^{ - 1}\!\! \left(\! {\frac{{r_1}}{\sqrt{r_{2}^2  + r_{3}^2 + r_{4}^2 }}}\! \right)\!,\,\theta _2  = \cot ^{ - 1}\!\! \left(\! {\frac{{r_2}}{\sqrt{r_{3}^2  + r_{4}^2 }}}\! \right)\!, \\ 
 \theta _3  = \cot ^{ - 1}\!\! \left(\! {\frac{{r_3}}{r_{4}}}\! \right)\!,\;\;\text{and}\;\;\phi _{1 \le k \le 4} \equiv \arg (c_{1 \le k \le 4}),  \\ 
 \end{array}
\label{eq:36}
\end{equation}
where $r_{k}\equiv |c_{k}|=|\langle{k}|\psi\rangle |$.  Then using (\ref{eq:36}) in (\ref{eq:34}) gives $(\theta _{4,1} ,\phi _{4,1} ) \equiv (\theta _1 ,\phi _1 )$, $(\theta _{4,2} ,\phi _{4,2} ) \equiv (\theta _2 ,\phi _2 )$, $(\theta _{4,3} ,\phi _{4,3} ) \equiv (\theta _3 ,\phi _3 )$, and $\chi _{4,3}  \equiv \phi _4 $, which then determine the essentially parameterized CK pairs $(a_{4,1} ,b_{4,1} )$, $(a_{4,2} ,b_{4,2} )$, and $(a_{4,3} ,b_{4,3} )$.  Using the abbreviations of (\ref{eq:14}), this gives us the \textit{determined} CK pairs as
\begin{equation}
\begin{array}{*{20}l}
   {(a,b)} &\!\! { \equiv (a_{4,1} ,b_{4,1} )} &\!\! { \equiv (\cos (\theta _1 )e^{i\phi _1 } ,\sin (\theta _1 ))}  \\
   {(c,d)} &\!\! { \equiv (a_{4,2} ,b_{4,2} )} &\!\! { \equiv (\cos (\theta _2 )e^{i\phi _2 } ,\sin (\theta _2 ))}  \\
   {(e,f)} &\!\! { \equiv (a_{4,3} ,b_{4,3} )} &\!\! { \equiv (\cos (\theta _3 )e^{i\phi _3 } ,\sin (\theta _3 )e^{i\phi _4 } )},  \\
\end{array}
\label{eq:37}
\end{equation}
where these $7$ angles are found in (\ref{eq:36}).  The \textit{free} CK pairs are then given in essential parameterization as
\begin{equation}
\begin{array}{*{20}l}
   {(g,h)} &\!\! { \equiv (a_{3,1} ,b_{3,1} )} &\!\! { \equiv (\cos (\theta _{3,1} )e^{i\phi _{3,1} } ,\sin (\theta _{3,1} ))}  \\
   {(j,k)} &\!\! { \equiv (a_{2,1} ,b_{2,1} )} &\!\! { \equiv (\cos (\theta _{2,1} )e^{i\phi _{2,1} } ,\sin (\theta _{2,1} )e^{i\chi _{2,1} } )}  \\
   {(l,m)} &\!\! { \equiv (a_{3,2} ,b_{3,2} )} &\!\! { \equiv (\cos (\theta _{3,2} )e^{i\phi _{3,2} } ,\sin (\theta _{3,2} )e^{i\chi _{3,2} } )},  \\
\end{array}
\label{eq:38}
\end{equation}
where the $8$ free parameters $\theta _{2,1} ,\theta _{3,1} ,\theta _{3,2}  \in [0,{\textstyle{\pi  \over 2}}]$ and $\phi _{2,1} ,\phi _{3,1} ,\phi _{3,2} ,\chi _{2,1} ,\chi _{3,2}  \in [0,2\pi )$ can have \textit{any} values.  Then, using (\ref{eq:37}) and (\ref{eq:38}) to generate $U^{[4]}$ from (\ref{eq:5}) as shown in (\ref{eq:15}), using notation similar to (\ref{eq:32}) we obtain a complete set of orthonormal state vectors, given by
\begin{equation}
\left\{ {|\psi \rangle ,|U_{2,:}^{[4]}\rangle ,|U_{3,:}^{[4]}\rangle ,|U_{4,:}^{[4]}\rangle} \right\},
\label{eq:39}
\end{equation}
which are the transposed row vectors of $U^{[4]}$ where the generating state is $|\psi \rangle =|U_{1,:}^{[4]}\rangle$.
\subsection{\label{subsec:IV-B}Multiport Interferometry}
As described in \cite[]{Reck}, a multiport interferometer (MPI) can be constructed by performing successive single qubit rotations on all possible pairs of spatially separated beams.  The resulting operation is a general unitary matrix $B$ of dimension $n$, the number of input ports.  

In such a setup, one uses a two-arm Mach-Zhender interferometer to implement each single qubit operation, with phase-shifter angles acting as the superposition and phase angles of each rotation.  As described earlier, if each qubit rotation is general, there are a large number of irrelevant parameters.  Therefore, using the essential parameterization method of (\ref{eq:26}) and (\ref{eq:28}), one can eliminate unnecessary hardware to achieve the simplest experimental setup.  Furthermore, by also using the first sector mapping of (\ref{eq:29}), one can minimize the angular ranges required from the superposition phase shifters.

Another benefit of these methods stems from the fact that the output spatial field modes of the MPI are
\begin{equation}
\hat{a}_{k}'\equiv B\hat{a}_{k}B^{\dag}  = \sum\limits_{m = 1}^n {V_{k,m} \hat{a}_m }=\sum\limits_{m = 1}^n {U_{m,k}^{*} \hat{a}_m },
\label{eq:40}
\end{equation}
where $\hat{a}_m$ is the annihilation operator for the \textit{input} field mode of location $m$, primes indicate output modes, and $V$ and $U$ are general unitary where $U\equiv e^{i\phi}U^{[n]}\equiv V^{\dag}$, and $U$ will simplify what follows. Note that $V$ and $U$ are not the same as the unitary $B$ of the MPI, which can be viewed as a multiport variable beam-splitter.  However, the transformation of (\ref{eq:40}) can be expressed directly in terms of phase shifter angles, and can be derived for a particular set of equipment using the methods of \cite[]{Reck}.

An example of a particular application is the explicit calculation of the output of an MPI given the input of a product of different coherent states.  By using (\ref{eq:40}) and the fact that coherent states are $|\alpha \rangle  = D(\alpha )|0\rangle$ where $D(\alpha ) \equiv e^{\alpha \hat{a}^{\dagger} - \alpha^{*} \hat{a}}$ is the displacement operator, we find
\begin{equation}
\begin{array}{*{20}c}
   {B\mathop  \otimes \limits_{k = 1}^n |\alpha _k \rangle } &\!\! { = \mathop  \otimes \limits_{k = 1}^n |\sum\nolimits_{m = 1}^n {U_{k,m} \alpha _m } \rangle  \equiv \mathop  \otimes \limits_{k = 1}^n |\alpha _k '\rangle },  \\
\end{array}
\label{eq:41}
\end{equation}
which yields the output as a product of coherent states, and also allows us to use the compact CK form of (\ref{eq:5}) to describe the new coherent state parameters (CSPs) $\alpha _k '$.

For example, given a $3$-input interferometer, with three coherent states $|\alpha \rangle$, $|\beta \rangle$, and $|\gamma \rangle$ in product as input, (\ref{eq:40}) and (\ref{eq:41}) reveal the MPI output to be
\begin{equation}
B|\alpha \rangle |\beta \rangle |\gamma \rangle  \equiv |\alpha '\rangle |\beta '\rangle |\gamma '\rangle ,
\label{eq:42}
\end{equation}
where, using $U\equiv e^{i\phi}U^{[3]}$ from (\ref{eq:13}), the new CSPs are
\begin{equation}
\begin{array}{*{20}l}
   {\alpha '} &\!\! { \equiv e^{i\phi } (a\alpha  + bc\beta  + bd\gamma )}  \\
   {\beta '} &\!\! { \equiv e^{i\phi } (b^* e\alpha  + ( - a^* ce - d^* f^* )\beta  + ( - a^* de + c^* f^* )\gamma )}  \\
   {\gamma '} &\!\! { \equiv e^{i\phi } (b^* f\alpha  + ( - a^* cf + d^* e^* )\beta  + ( - a^* df - c^* e^* )\gamma )}.  \\
\end{array}
\label{eq:43}
\end{equation}
Thus, plugging the essential parameterization of (\ref{eq:28}) into (\ref{eq:43}), with the help of (\ref{eq:12}), allows us to determine precisely how the phase shifters control the coherent state parameters in the output state of (\ref{eq:42}).

From (\ref{eq:43}), the sum of square magnitudes of the CSPs is invariant; $|\alpha '|^2  + |\beta '|^2  + |\gamma '|^2 =|\alpha |^2  + |\beta |^2  + |\gamma |^2 $, so
\begin{equation}
\sum\nolimits_{k = 1}^n {|\alpha _k '|^2  = \sum\nolimits_{k = 1}^n {|\alpha _k |^2 } },
\label{eq:44}
\end{equation}
which is because the vectors of $U$ are orthonormal.

A \textit{physical} application of all this is the ability to calibrate a multiport interferometer with classical inputs.  By setting different combinations of the coherent state inputs to vacuum in (\ref{eq:42}), one can isolate any problems in the output of a large MPI to only a few possible culprit qubit rotations.  Then, once the proper $U$ is verified using the convenient classical calibration beams, one can remove them and proceed with any other input desired.
\subsection{\label{subsec:IV-C}Entanglement Preserving Transformations}
The methods of this paper are also useful in finding explicit forms for entanglement preserving transformations.  For $N$-party systems, these are typically generated as
\begin{equation}
U_{EP}  \equiv e^{i\gamma} e^{ - i{\textstyle{{\alpha _1 } \over 2}}\mathbf{u}_{(1)}  \cdot \bm{\lambda} ^{[n_{1} ]} }  \otimes  \cdots  \otimes e^{ - i{\textstyle{{\alpha _N } \over 2}}\mathbf{u}_{(N)}  \cdot \bm{\lambda} ^{[n_{N} ]} },
\label{eq:45}
\end{equation}
where $\gamma$ and $\alpha_m$ are angles, the $\mathbf{u}_{(m)}$ are real unit vectors of dimension $n_{m}$, and $\bm{\lambda}^{[n_{m}]}$ are vectors of generalized Gell-Mann matrices of dimension $n_{m}$.  However, matrix exponentiation can be avoided completely if we use (\ref{eq:5}) with the essential parameterization of (\ref{eq:28}) to write
\begin{equation}
U_{EP}  = e^{i\gamma} U_{k_1 }^{[n_1 ]}  \otimes  \cdots  \otimes U_{k_N }^{[n_N ]},
\label{eq:46}
\end{equation}
where subscripts $k_m$ indicate that each unitary matrix has a separate set of essential parameters, resulting in $D_{EP}  =1- N+ \sum\nolimits_{m = 1}^N {n_m^2 }$ variables, the same as in (\ref{eq:45}).  

For example, in a qubit-qutrit system, if $\gamma=0$ in (\ref{eq:46}),
\begin{equation}
\begin{array}{*{20}l}
   {U_{EP}  = U_{1}^{[2]}\otimes U_{2}^{[3]}=}  \\
   {\left(\!\!\! {\begin{array}{*{20}c}
   {a_1 } & {b_1 }  \\
   { - b_1^* } & {a_1^* }  \\
\end{array}}\!\! \right)\! \otimes\! \left(\!\! {\begin{array}{*{20}c}
   {a_2 } & {b_2 c_2 } & {b_2 d_2 }  \\
   {b_2^* e_2 } & { - a_2^* c_2 e_2  - d_2^* f_2^* } & { - a_2^* d_2 e_2  + c_2^* f_2^* }  \\
   {b_2^* f_2 } & { - a_2^* c_2 f_2  + d_2^* e_2^* } & { - a_2^* d_2 f_2  - c_2^* e_2^* }  \\
\end{array}}\!\! \right)\!,}  \\
\end{array}
\label{eq:47}
\end{equation}
which is subject to (\ref{eq:26}) and (\ref{eq:28}).  Thus, the essential parameterization lets us express entanglement preserving transformations without the need for exponentiation.
\section{\label{sec:V}CONCLUSIONS}
The methods presented here produce unitary matrices in an orderly, predictable fashion, highlighting the connection between unitary matrices, pure states, and hyperspherical coordinates.  These conventions reveal the true structure of unitary matrices with constraints already built-in, simplifying tasks such as generation of orthonormal bases, multiport interferometry, and generating entanglement preserving transformations.  The methods of this paper may also be useful for differentiation, optimization, approximation, and quantum control.

The presented conventions are designed to be convenient for many applications in quantum mechanics, but they are not unique.  Over the years, several methods have been developed to parameterize unitary matrices, such as in \cite[]{Dit1,Tilm,Dit2,Jarl,Fuji,Spen}, and some methods may be more useful than others depending on the application.

The most general unitary matrix, as expressed using the conventions of this paper, is $G^{[n]}\equiv e^{i\gamma}U^{[n]}$, where $\gamma\in[0,2\pi)$, and $U^{[n]}$ is given by (\ref{eq:5}) and parameterized by (\ref{eq:26}) subjected to (\ref{eq:28}) to ensure that $G^{[n]}$ has exactly $n^2$ variables.  Thus, this method allows us to construct the most general unitary matrices using the fewest parameters possible \textit{without} the need for exponentiating combinations of generators of $\text{U}(n)$, producing a simple form that is both predictable, and readily applicable to a wide variety of fields of interest in physics.
\begin{acknowledgments}
Many thanks to Ting Yu for his guidance, patience, and helpful feedback.  This work was supported by the I\&E Fellowship at Stevens Institute of Technology.
\end{acknowledgments}
%
\end{document}